\documentclass[preprint]{aastex}
\usepackage{epsfig}
\begin{document}

\title{Spatially Resolved Millimeter Spectroscopy of the Gravitational
Lens PKS 1830-211}
\author{Jonathan J. Swift \& William J. Welch}
\affil{Radio Astronomy Laboratory, and Department of Astronomy,
University of California, Berkeley, CA 94720}
\author{Brenda L. Frye}
\affil{Lawrence Berkeley National Laboratory, Department of Physics,
University of California, Berkeley, CA 94720}

\begin{abstract}

This paper presents data from the BIMA interferometer showing
spatially resolved absorption spectra of the gravitationally lensed
quasar PKS 1830-211. High-resolution (1.2\,km\,s$^{-1}$) spectra were
taken in two spectral windows centered on the redshifted frequencies
of the HCO$^+(2\leftarrow 1)$ and HCN$(2\leftarrow 1)$ molecular
transitions. There is no molecular absorption in the northeast image
but the southwest image reveals optically thick absorbing gas at these
transition frequencies. Further analyses conclude that the spectra are
consistent with completely saturated absorption in the southwest image
and the line profiles suggest that the absorbing medium is complex,
perhaps containing multiple components and small scale structure. The
absorption occurs along a pencil beam through the lensing galaxy which
is thought to be a late type spiral oriented almost face on. However,
the spectra show absorption spanning more than 60\,km\,s$^{-1}$ which
is difficult to explain for this scenario. 

\end{abstract}


\section{Introduction}

PKS 1830-211 is a radio bright source determined to be a
gravitationally lensed quasar from its ring-like morphology
\citep{jau91,sub90,rao88} toward which molecular absorption has been
detected \citep{wc96}. The object consists of two flat spectrum images
connected by a nearly circular, low surface brightness steep spectrum
Einstein ring \citep{jau91,koc92}. The two bright sources are
separated by  $\sim 1''$ and lie to the northeast (NE) and southwest
(SW) of the ring center.

Wiklind and Combes (1996) discovered twelve different rotational
transitions involving five different molecular species toward PKS
1830-211 in unresolved spectra. The lines had widths of $\sim
30$\,km\,s$^{-1}$, depths of $\sim 40$\%, and were redshifted to
$z=0.89$. Based on standard isotopic ratios, the detection of
H$^{13}$CO$^+(2\leftarrow1)$ implied that the HCO$^+(2\leftarrow1)$
was optically thick. This assumption along with published flux ratios
of the NE and SW components led to the conclusion that the absorption
was only in the southwest component. Spectra of the two main
components were obtained with the Berkeley Illinois Maryland
Association (BIMA) array by resolving the source in the 3-mm band
\citep{fry96}. The molecular absorption only appeared in the SW
component as was expected, but did not reach the zero flux
level. These results seemed to conflict with an analysis done by
Wiklind and Combes (1998) on the unresolved spectrum of PKS 1830-211
where the shift in the field phase center at the frequency
corresponding to the HCO$^+(2\leftarrow1)$ transition was consistent
with saturated absorption in the SW image. 

High-resolution HCO$^+(2\leftarrow1)$ spectra of the unresolved
source taken with the Plateau de Bure interferometer \citep{wc98}
showed that the trough of the absorption is not flat as would be
expected from complete saturation. The line profile was also clearly
asymmetric with the broader wing toward  negative velocities. These
results are reasonable for absorption due to a multi-component cloud
with possible small scale structure rather than one optically thick
cloud.  

Galactic models \citep{wc98}, a determination of the quasar redshift,
$z=2.5$ \citep{lid98}, and lensing models \citep{nai93,koc92} combined
to form a coherent picture of this gravitational lens system and the
Hubble parameter was determined to be $H_0 =
65^{+15}_{-9}$\,km\,s$^{-1}$\,Mpc$^{-1}$ \citep{lid98}. The rather
large uncertainty in the time delay measurement, about 20\%
\citep{lov98}, is the major source of error in this estimate of
$H_0$. Continued monitoring of the flux ratio of PKS 1830-211 would
allow for a better estimate of the time delay and therefore a better
determination of the Hubble parameter. The flux ratio can be easily
determined from an unresolved spectrum of the source if the absorption
in the SW image is saturated \citep{wc96}. While many lines have been
studied in this system, high-resolution spectra of saturated
absorption in the resolved SW image have yet to be presented.

This paper presents observations of PKS 1830-211 made with the BIMA
interferometer in Hat Creek, California \citep{wel96} which yielded
spectra of the two individual components. There have been a couple of
important improvements to the BIMA array since the published results
of Frye et al. (1996). The most significant upgrade has been the
installation of new SIS mixers in 1997 which has halved the receiver
noise temperatures \citep{eng98}. There have also been new A-array
stations added which produce a better imaging beam. The observations
presented here were done with higher spectral resolution than the
previous BIMA observations and were intended to give a more detailed
look at the line profile of the resolved SW image. In {\S} 2 the
observations and observing conditions are outlined and the details of
the self-calibration procedure used in the data reduction are
given. Section 3 presents flux values and an analysis of the
spectra. Rotation temperatures, column depths, and the implications of
the line profiles are then discussed in {\S} 4.

\section{Observations and Data Reduction}

PKS 1830-211 was observed in the 3-mm band using the ``A-''
configuration of the BIMA array which is the standard A configuration
without the longest baselines at 1.9\,km. The correlator was set so
that in each sideband there were two spectral windows containing 128
channels with a spectral resolution of 1.2\,km\,s$^{-1}$ and four wide
band windows each containing 32 channels and spanning 100\,MHz. The
first local oscillator was tuned so that the redshifted
($2\leftarrow1$) rotational transitions of HCO$^+$ and HCN, at
94.588\,GHz and 93.977\,GHz respectively, fell into the two high
resolution windows. The HCO$^+$ line was Doppler tracked throughout
the observations. The standard quasar 3C273 was observed to calibrate
the passband, Uranus gave an absolute flux scale, and 1733-130 was
used as a secondary calibrator. PKS 1830-211 was observed in 30 minute
intervals using 23 second integrations with the intention of self
calibration. 

Data were used from two separate observing tracks on 1999 December 27
and 28. The observations on the 27th were good with typical system
temperatures between 200\,K and 300\,K and the RMS phase variations
measured on 100\,m baselines were on the order of 200\,$\mu$m. The
atmospheric conditions on the 28th were even better with slightly
lower system temperatures and phase variations of $\sim 80\,\mu$m.

PKS 1830-211 is a bright source which can be self calibrated on a
record-by-record basis. The data on each day was split into two datasets
corresponding to the upper and lower side bands and the
phases were first corrected in the lower side band using an iterative
self-calibration process. The first iteration used a point source at
the observing center as a model, and produced a two-component
image. The inverted CLEAN model containing two main components was then
used as the self-calibration model in the subsequent iteration. Two
self-calibration iterations were needed for the image to converge. The
upper side band was then phase calibrated in the same way with the
only difference being that the first model used in the
self-calibration iteration was the CLEAN model from the lower side
band. The spectral windows were then phase calibrated channel by
channel using a 65 minute time interval allowing enough signal for a
reliable calibration. The bandpass was determined from 3C273 and was
applied to the source and then the datasets were scaled to the flux of
Uranus measured on December 27. The calibrated datasets were then
combined and inverted into a single plane upper side band continuum
map and two 128 plane spectral cubes. All reductions were done using
the MIRIAD software package.   

The two main components were not resolved to a sufficient degree in
all of the data due to the position angle of the synthesized beam and
therefore some data were discarded. A local sidereal time range of
1700--1900 was used on both nights covering an elevation range from 17
to 27$^{\circ}$. The resulting synthesized beam dimensions are
$1\arcsec.08 \times 0\arcsec.59$ FWHM with a position angle of
$-11.6^{\circ}$. This resolution issue and the implications will be
considered further in the following section.

\section{Analysis and Results}

The combined datasets yield a NE continuum level of $1.13\pm 
0.1$\,Jy\,beam$^{-1}$ and a SW level of $0.96\pm 0.1$\,Jy\,beam$^{-1}$
with the flux levels of the individual datasets agreeing to within
10\%. Although this is known to be a highly variable source
\citep{van95}, these levels are reasonable in comparison to previous
results \citep{wc98,fry96,wc96}. The flux ratio of the two images,
$1.18\pm 0.06$, is expected to be more accurate than the overall flux
scale and agree to within 3\% on the two days. The separation angle
was determined using the inverted CLEAN model of the combined datasets
giving a value of $0\arcsec.99\pm0\arcsec.05$, consistent with
previously published values \citep{rao88,sub90,nai93,joh96,fry96}.

Spectra were taken from the two 128 plane data cubes containing the
HCO$^+(2\leftarrow 1)$ and HCN$(2\leftarrow 1)$ transitions at the
locations of the continuum peaks (see Fig. \ref{f1}). The line
profiles obtained here are in good qualitative agreement with the high
resolution (0.5\,km\,s$^{-1}$) spectrum of HCO$^+(2\leftarrow 1)$ in
Fig. 7 of Wiklind and Combes, 1998 (hereafter 7WC). The
absorption widths are comparable at about 45\,km\,s$^{-1}$ FWHM, and
the asymmetry in the lines is apparent with the broader side toward
negative velocities. There is some residual flux in the trough of the
absorption, however there is no apparent slope as seen in 7WC. 

The absorption in the SW image reaches the zero level with 5 channels
at or below zero flux in the HCO$^+$ window and 3 channels in the HCN
window. In reference to 7WC, a numerical average over
20\,km\,s$^{-1}$ 
(16 channels) around the centroid of the lines was computed giving an
average flux level in the absorption trough. This procedure applied to
the high-resolution profile of HCO$^+(2\leftarrow 1)$ in 7WC
yielded a value of $\sim$ 0.04--0.06\,Jy due to the slope of flux in
the absorption trough assuming that the negative velocity edge of the
the narrow profile corresponds to a zero flux level in the SW image. 
The NE and SW components are known to have sizes at or below the
milliarcsecond scale \citep{joh96}, however the two images in our
dataset did not deconvolve into two point sources. A no noise, no phase
error model of our observations revealed flux contamination between
the two images at the level of 0.01\,Jy\,beam$^{-1}$ preventing a double point
source deconvolution. Therefore the flux of a saturated line averaged
across the narrow trough given the profile of 7WC and our resolution
determined from models is expected to be between
0.05--0.07\,Jy\,beam$^{-1}$. The average flux levels measured in the
absorption troughs of the present data are 0.058 and
0.066\,Jy\,beam$^{-1}$ for the HCO$^+$ and HCN lines respectively. 

The data show no evidence of absorption in the NE component in HCO$^+$
however a small absorption feature is noticeable in the HCN window at
0\,km\,s$^{-1}$. This feature appeared only in the December 28 data
and is small but statistically significant reaching the $3\sigma$
level. It is interpreted as a small amount of contamination
from the SW component. 

\section{Discussion}

This dataset is unique in showing direct evidence of spatially and
spectrally resolved saturated absorption in PKS 1830-211. The noise in
our spectra, with a measured RMS of 0.098\,Jy\,beam$^{-1}$, is too
high to verify the upward slope from negative to positive velocities
in the HCO$^+$ window which is seen in 7WC. However, the absorption
does reach zero flux levels in multiple channels in each window and
the profile is consistent with saturation given the profile in 7WC and
the observational models constructed.

These results are contrary to the results obtained by Frye et
al. (1996) who also used the BIMA array to resolve the two components,
but obtained absorption spectra that were only $\sim 80\%$
saturated. The data quality and the judiciousness in setting the
self-calibration parameters determine the reliability of an iterative
self-calibration process. Lower data quality leads to longer, more
elaborate calibrations which often require extensive flagging and
ultimately challenge the soundness of the result. Starting with a
self-calibration model that doesn't represent the known spatial
distribution of flux only makes the process more unstable. Our data
were taken in better conditions than the previous BIMA observations
and with new SIS receivers, both of which contributing to a higher
overall data quality. There were also different A-array antenna
stations producing a better beam from which this image based reduction
may have benefited. The calibration process involved only two
iterations and varying the self-calibration parameters, such as the
calibration interval and convergence criteria, produced consistent
results. All of these considerations give us much confidence in the
robustness of our result.

The residual flux in the absorption trough is thought to be due to
small scale structure in the absorbing cloud which translates into a
varying covering factor as a function of velocity
\citep{wc98,wc96,fry96}. The asymmetry seen in the lines is also
evidence of a complex, inhomogeneous absorbing medium. However, in
order to get any quantitative information out of these profiles it is
necessary to make simplifying assumptions about the equilibrium state
of the gas and the structure.

The deep absorption prevents a direct determination of the
physical parameters of the absorbing medium, however limits can be
determined using a lower limit of the velocity integrated optical
depth. The HCO$^+(2\leftarrow 1)$ and HCN$(2\leftarrow 1)$ line
profiles presented here with their $(1\leftarrow 0)$ counterparts
presented in Menten et al. (1999) and Carilli et al. (1998) give
low values for the rotation temperatures, $\lesssim 6$\,K, consistent
with previous analyses \citep{wc96,men99}. This result implies that
HCN and HCO$^+$ are primarily excited by the CMB, so we can assume
that the rotation temperature of these molecules are $T_{rot}\sim
5$\,K which is the temperature of the CMB at $z=0.89$. This value of
$T_{rot}$ along with the calculated velocity integrated optical depths
give lower limits to the column densities for HCN and HCO$^+$;
$N($HCN$) \gtrsim 3\times10^{14}$cm\,$^{-2}$ and $N($HCO$^+) \gtrsim
2\times10^{14}$\,cm$^{-2}$.

The lensing galaxy in which this absorption is taking place is likely
a massive, early-type spiral \citep{wc98,wc96}. Given that the
abundance ratios are somewhat similar to the Milky Way, the column
densities derived above imply an H$_2$ column of $N($H$_2)\gtrsim
10^{22}$ but the low average excitation temperatures of the HCN and
HCO$^+$ lines imply a moderately low average volume density ($\lesssim
10^{4} $cm$^{-3}$). Therefore the pathlength through the cloud must be
greater than 1\,pc, but it is not clear how extended the cloud is in
the line of sight direction. The full velocity range over which there
is absorption is $> 60$\,km\,s$^{-1}$ which is not atypical for
observations in the Milky Way \citep{fry96}. However according to
galactic \citep{wc98} and lensing \citep{nai93,koc92} models, the
lensing galaxy of PKS 1830-211 is oriented almost face on ($i\simeq
16^{\circ}$)  and the light from the SW image is traversing the
lensing galaxy at a galactocentric distance between 1.8\,kpc and
3\,kpc. Assuming this orientation, it is hard to imagine how the
absorption occurs over such a large velocity range (see Menten et
al. 1999).

\section{Summary}

The two main components of the gravitational lens PKS 1830-211 have
been resolved with the BIMA interferometer. The continuum levels are
$1.13\pm 0.1$\,Jy\,beam$^{-1}$ for the NE component and $0.96\pm
0.1$\,Jy\,beam$^{-1}$ for the SW component giving a flux ratio of
$1.18\pm 0.06$. These levels are expected to vary, but are consistent
with other published values. 

High-resolution spectra (1.2\,km\,s$^{-1}$) of the redshifted lines of
HCO$^+(2\leftarrow 1)$ and HCN$(2\leftarrow 1)$ were shown to reach
the zero flux level in the SW image but with a non zero flux level
averaged across the narrow trough. These results are an improvement
over the previous BIMA observations of this object \citep{fry96} and
are in agreement with what would be expected from saturated absorption
by a complex absorbing cloud with multiple components and small scale
structure.

Comparing our line profiles with previously published spectra of the
HCO$^+(1\leftarrow 0)$ and HCN$(1\leftarrow 0)$ transitions in the SW
component \citep{men99,car98} gives a low upper bound on the
excitation temperature, $\lesssim 6$\,K, which translates to fairly
high column densities, $N($HCN$) \gtrsim 3\times10^{14}$\,cm$^{-2}$
and $N($HCO$^+) \gtrsim 2\times10^{14}$\,cm$^{-2}$. The high column
densities imply a large H$_2$ column but the low excitation
temperatures suggests that the volume density of H$_2$ is moderately
low. Given the estimated galactocentric distance of the absorbing
cloud $\geq 1.8$\,kpc and the inclination angle of the galaxy, $i
\simeq 16^{\circ}$, the line widths are remarkably large.

\bigskip

We would like to thank Dick Plambeck for helping with the many
questions that arose during this research. Also thanks to Mel Wright
for help with generating model observations and for general
advice. Michiel Hogerheijde and Geoff Marcy both made helpful
contributions. BLF would like to thank Joe Silk and Tom Broadhurst for
their ideas and contributions. This research has been funded by NSF
grant AST-9981308.

\clearpage

\figcaption[figure1.eps]{Upper sideband continuum map with the spectra
from the NE and SW components overlayed. The contour levels are 
10, 20, 30, 40, 50, 60, 70, 80, and 90\%. \label{f1}}

\clearpage
\begin{figure}[ht]
\begin{center}
\rotatebox{270}{{\epsfig{file=figure1.eps,width=6.5in}}}
\end{center}
\end{figure}

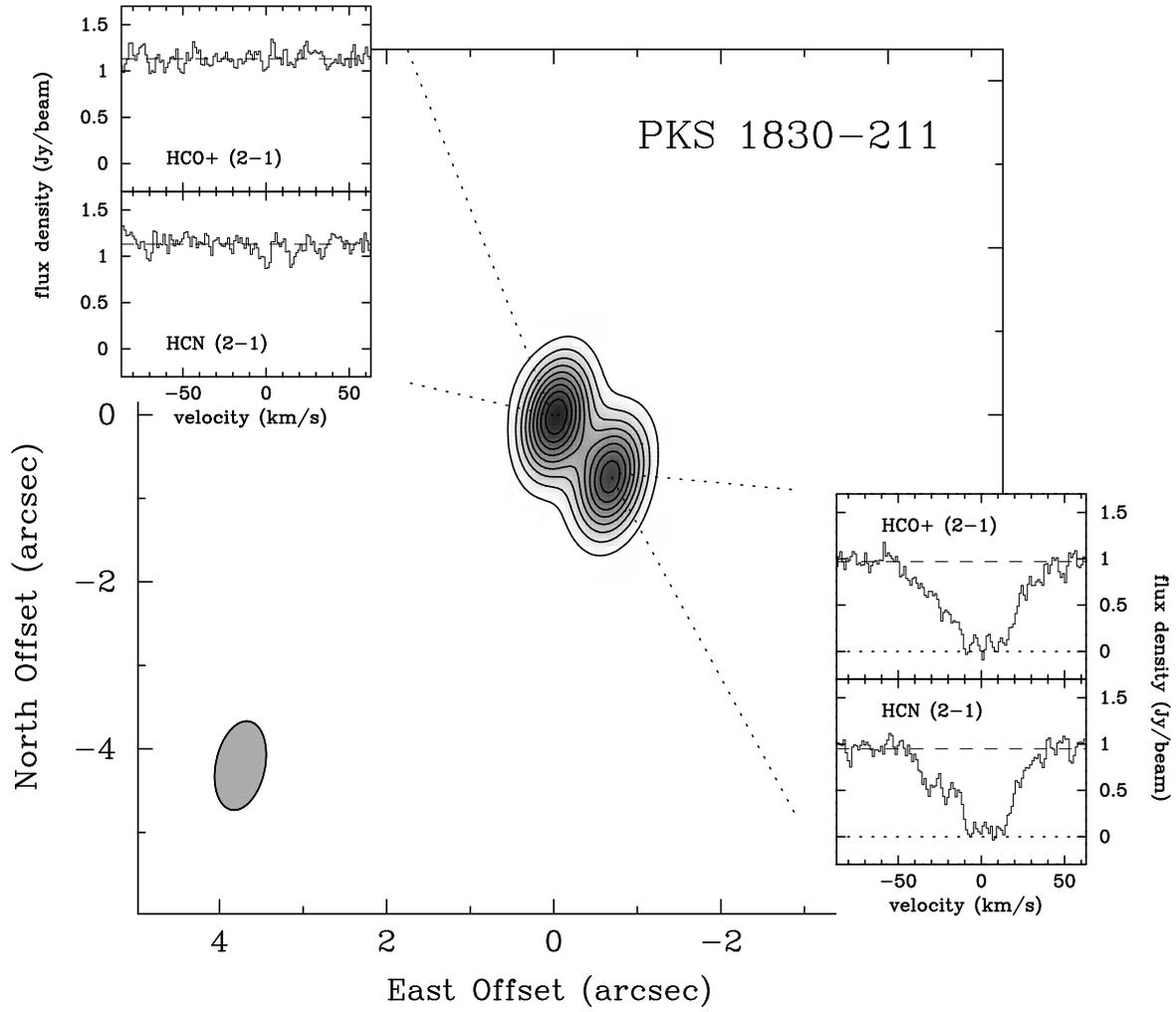
\end{document}